# Design and Implementation of a Low-Latency and High-Reliability System Based on Software-Defined Radio (SDR)

Zhiwen Wu
Intelligent Computing and Communication Lab
Beijing University of Posts and Telecommunications
Beijing, China
e-mail: sswzw@bupt.edu.cn

Tianpeng Yang
Intelligent Computing and Communication Lab
Beijing University of Posts and Telecommunications
Beijing, China
e-mail: ytpbupt@163.com

Haitao Liu
Technology Innovation Center
China Telecom Research Institute
Beijing, China
e-mail: liuht.bri@chinatelecom.cn

Geng Shi
Intelligent Computing and Communication Lab
Beijing University of Posts and Telecommunications
Beijing, China
e-mail: shigengbupt@163.com

Kan Zheng
Intelligent Computing and Communication Lab
Beijing University of Posts and Telecommunications
Beijing, China
e-mail: zkan@bupt.edu.cn

*Abstract*—Ultra-reliable and low-latency communication (URLLC) is one of the three major service classes supported by the fifth generation (5G) New Radio (NR) technical specifications. In this paper, we introduce a physical layer architecture that can meet the low-latency and high-reliability requirements. The downlink system is designed according to the Third Generation Partner Project (3GPP) specifications based on software defined radio (SDR) system. The URLLC system physical layer downlink is implemented on the open source OpenAirInterface (OAI) platform to evaluate the latency and reliability performance of the scheme. Not only the URLLC system reliability performance is tested based on the simulation platform, but also the delay performance is evaluated by the realization of the over-the-air system. The experimental results show that the designed scheme can approximately meet the reliability and delay performance requirements of the 3GPP specifications.

## I. INTRODUCTION

The fifth generation (5G) provides a large number of services, e.g., ultra-reliable and low-latency communication (URLLC) with new features of high-reliability and low-latency [1]. Thanks to the features, URLLC is widely considered a key technology in multiple scenarios including industrial automation [2], smart grids [3] and vehicle safety [4]. Diversified quality of service requirements place higher demands on wireless networks [5]. At the beginning of 2016, the Third Generation Partner Project (3GPP) initiated the development of 5G protocols. The R16 version of the protocols including the standardization of URLLC is scheduled to be completed by the end of 2019. The 3GPP conference has set a range of standards and performance requirements, e.g., the cases require the end-to-end delay to be in milliseconds and the reliability to be no less than 99.999% for a packet of 32 bytes in no more than 1 ms latency [4]. Higher requirements are also imposed on other aspects of wireless communication performance, e.g., energy efficiency [6]. In this paper, a shorter Transmission Time Interval (TTI) and a robust coding scheme are introduced to implement a low-latency and high-reliability system.

So far, researchers have proposed many experimental platforms for the URLLC system, but most of them are for simulation. To carry out real-time evaluation performance testing, this paper realizes the software defined radio (SDR) platform for URLLC according to the 3GPP specifications. The SDR system, using programming languages to implement traditional communication systems based on universal processors, can be effectively completed and upgraded because of advanced programming languages and existing libraries [8][9][10]. At present, there are many sophisticated SDR platforms, e.g., OpenBTS [11], srsLTE [12] and Amarisoft [9][13]. OpenAirInferface (OAI), an open source platform, is one of the most comprehensive and widely used platforms [14][15]. It follows up the New Radio (NR) specifications in time and provides the communication module programs. Compared with other software platforms, OAI has great advantages of openness and flexibility in testing and validating the new technologies of communication [16].

This paper makes modifications to the physical layer in OAI to implement the URLLC system, e.g., TTI, reference signal (RS) and data processes of the Physical Downlink

Shared Channel (PDSCH). Other parts are also modified including secondary synchronization signal (SSS) and primary synchronization signal (PSS). The contributions of this paper are summarized as follows, i.e.,

1) The physical layer downlink system is well designed and implemented for the optimization of reliability and latency performance based on the OAI.

2) A series of experiments of the OAI platform are conducted to evaluate the physical layer downlink reliability and latency performance.

The structure of the paper is as follows, Section II briefly introduces the physical layer technical architecture. In Section III, the system is introduced including the design and implementation of TTI, the location setting of RS, and PDSCH processes. The experimental results of the reliability and delay test based on the platform are presented in Section IV. Finally, concluding remarks are drawn in Section V.

## II. SYSTEM ARCHITECTURE

A GPP and a radio transceiver peripheral can form the evolved Node B (eNB) or the User Equipment (UE) of the SDR system [8]. The communication program running on the GPP processes the baseband signal and sends the data to the radio transceiver peripheral through the USB3.0 interface. The radio transceiver peripherals are responsible for the frequency modulation.

Based on the OAI platform, the physical layer downlink of the URLLC system is implemented. Some physical layer processes are as shown in Fig. 1. The GPPs and the radio transceiver peripherals are also shown. Firstly, the eNB physical layer processes the control information, including the Physical Control Format Indicator Channel (PCFICH) information, the DCI, the Physical Downlink Control Channel (PDCCH) scrambling, PDCCH interleaving and PDCCH mapping shown as Fig. 2. Secondly, the PDSCH and RS are processed including PDSCH encoding, scrambling, modulation, generation of RS and orthogonal frequency division multiplexing (OFDM). The PDSCH encoding can be subdivided into Cyclic Redundancy Check (CRC) addition, segmentation, low-density parity-check (LDPC) encoding, and rate-matching modules. The LDPC scheme is following the 5G protocols [13]. The above processes are carried out on the eNB GPP. Then the GPP sends the baseband data to the radio transceiver peripheral. The peripheral modulates the baseband signal to the high frequency and completes the delivery.

The radio receiver device demodulates the signal from high frequency to low frequency and delivers the data to the UE GPP. Then the physical layer of the UE processes the baseband signal including channel estimation, control information decoding, and PDSCH decoding. Finally, the reliability and delay performance of the URLLC system is evaluated by the error block rate and the physical layer delay.

Other physical layer processes of the URLLC system, e.g., broadcast signals, still follow the LTE protocols. To meet the requirements of the low-latency performance, the scheduling unit of physical layer downlink is no longer the sub-frame in the LTE. The introduction of the new concept mini-slot provides feasibility for the system implementation.

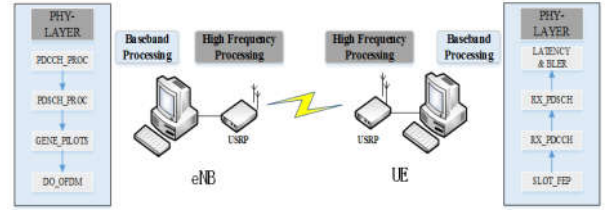

Figure 1. Downlink physical layer architecture of the system

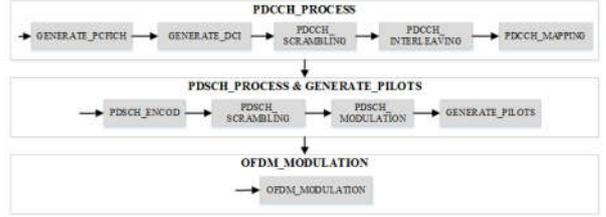

Figure 2. Physical layer process of the eNB

## III. DESIGN AND IMPLEMENTATION OF THE DOWNLINK PHYSICAL LAYER SYSTEM

The downlink physical layer system achieves the target of the low delay performance based on the configuration of the scheduling unit. In terms of reliability performance, the system is implemented based on the PDSCH code scheme of 3GPP specifications. It includes the flexible and variable Modulation and Coding Scheme (MCS), which is determined by the Channel Quality Indicator (CQI). The configuration of RS and control information also needs to be designed and implemented for the system reliability performance.

### A. The physical layer technology for the optimization of downlink delay performance

On the eNB side of the over-the-air platform, the scheduling unit is the mini-slot. The platform over the air requires time synchronization between the eNB and the UE. First slot (Slot0) and sixth slot (slot5) both contain seven symbols, of which the last two symbols are used to place PSS and SSS. Other slots contain four symbols in the system. Since the time duration of the two kinds of mini-slots is different, they need to be processed separately. After the synchronization signal is extracted from the longer slot, the two kinds of slots are handled in the same way. Besides, since the eNB and UE of the simulation platform are running on the same computer, there is no synchronization signal in mini-slot. Each mini-slot contains four symbols.

Compared with the LTE protocols, the difference of the NR specifications is mainly focused on the physical layer structure. Therefore, this paper bases on the SDR physical layer downlink to implement the URLLC system. To realize low-latency performance, a shorter TTI is required. The duration of a TTI is determined by the symbol number and the duration of per symbol.

A downlink mini-slot can contain 2/4/7 symbols [7]. According to the principle of communication, the duration of a symbol decreases with the subcarrier spacing increasing. Decreasing the duration of per symbol and using fewer symbols in a mini-slot can both meet the requirements of low-latency performance. Since the experiment takes full account of the characteristic of OAI, the radio transceiver peripherals in use and resource utilization, the system uses 15 kHz as the subcarrier spacing.

Due to the limitations of the USRP, the scheduling unit of the experimental platform is modified as shown in Fig. 3. When the eNB transmits data, 14 symbols are used as the basic transmission unit. The data to be transmitted are placed in the first four symbols, and the last ten symbols are in an idle state. The eNB transmits the data and the UE receives the data at the same time. After the four symbols are received, the UE processes the received data contained in the first four symbols. The data in the next ten symbols are discarded by the UE.

*B. The physical layer technology for improving downlink reliability performance*

*1) RS and control information:* The placement of RS is determined according to the 3GPP specifications [17]. Only the first symbol (Symb0) and the fourth symbol (Symb3) contains RS in the URLLC system. Because the subcarrier spacing of 15 kHz makes the channel correlation between continuous subcarriers stronger, more data can be placed between adjacent reference signals in the system. The adjacent reference signals use six subcarriers as the spacing of continuous RS in the system. The scheme can improve resource utilization. The first RE is as the start place for the RS in Symb0 and the RS starts from the forth RE in Symb3. The scheme of the RB is as shown in Fig. 4.

That the UE uses RS to implement the channel estimation can be summarized as follows three steps in the experiment. (i) The channel state at the RS position is estimated by using the least-square (LS) method. (ii) The channel state without RS in Symb0 and Symb3 is estimated by using the linear filter interpolation algorithm. (iii) The same algorithm with step 2 is used to estimate the channel state of the second symbol (Symb1) and the third symbol (Symb2) by using the state of Symb0 and Symb3. Then the data are extracted from the RB based on these channel conditions.

In the URLLC system, the data of PCFICH, PHICH and PDCCH are completely placed in Symb0. Control Format Indicator (CFI) is placed evenly on the frequency domain in Symb0. The PCFICH occupies four resource element groups (REG) and each REG is distributed in RB0/6/12/18 for 5 MHz bandwidth. Then the resource for PHICH is distributed. Fig. 4 shows the distributed resources for PCFICH and PHICH of the system. Resources not used for RS in symb1, symb2 and symb3 are used to transmit the data in PDSCH.

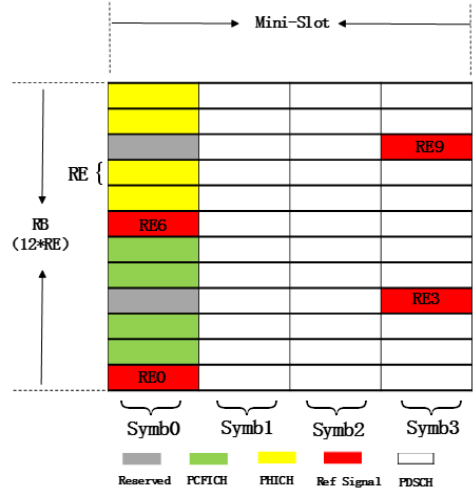

Figure 4. RB of the URLLC system

In the simulation platform, since the eNB and the UE run on the same computer, the source data of the control information, e.g., the DCI generated by the eNB, are placed in the GPP memory for the UE to extract shown as Fig. 5. After the UE completes the channel estimation steps, the data of the PDSCH are extracted from the resource based on the control information, which is directly got from the memory. This scheme eliminates the impact of inaccurate control information on the reliability of PDSCH. Since the over-the-air platform eNB and the UE run on different computers, the control information and the PDSCH data both need to be extracted after the channel estimation is completed.

*2) MCS and PDSCH:* As shown in Fig. 4, the resources available for PDSCH have been determined in 3GPP R15. The Transport Block Size (TBS) for the URLLC system can be determined by (1) combined with other specifications in 3GPP R15 [18].

$$N_{info} = N_{RE} * R * Q_m * v \quad (1)$$

where $N_{info}$ refers to the bit number of one unit. $N_{RE}$ refers to the number of resource elements (RE) available for PDSCH and $R$ stands for the channel bit rate. $Q_m$ refers to the modulation order used, and $v$ refers to the number of system layers. Since it does not conform to the high-reliability characteristic of the URLLC system when the modulation order is higher, the modulation order in the experiment is limited to two. The bandwidth is 5 MHz in the system. The specific information is as shown in Table I [18].

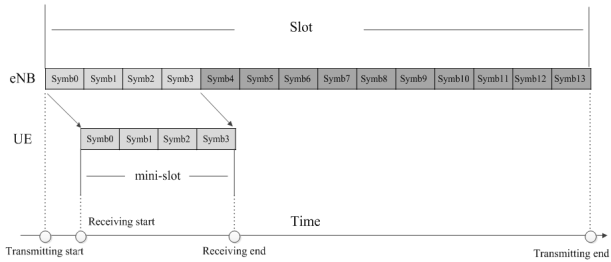

Figure 3. Transmitting and receiving data process

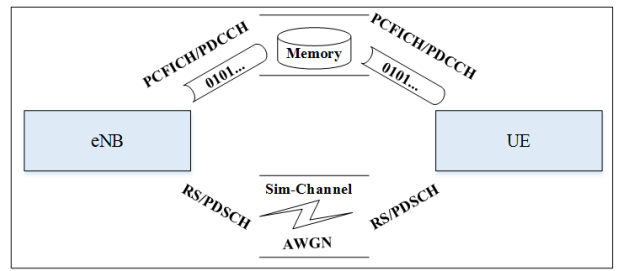

Figure 5. Data transmission process of simulation platform

As is known from Table I, TBS is smaller which does not need to be segmented specified when four symbols are used in a mini-slot for scheduling and QPSK is used as the modulation mode. The BG2 is used as the base matrix of LDPC encoding and decoding [19]. Besides, the CRC addition, rate-matching and other modules related to the encoding and decoding are implemented in the system [13].

To maintain the stability of the OAI platform, the URLLC system uses two threads to process the odd slot and the even slot separately as shown in Fig. 6. When the GPP receives the data from the PDSCH, the software of the UE calls the corresponding thread according to the received slot number, and then the data are processed by the segmentation module, rate-matching module, etc. until the previous step of decoding module. The LDPC decoding module of the over-the-air platform is accelerated by using the Advanced Vector Extension (AVX) instruction set to reduce the physical layer downlink delay of the URLLC system [13]. The simulation platform still decodes the data using a normal LDPC decoding module. Therefore, appropriate decoding modules need to be selected for different platforms as shown in Fig. 6.

## IV. EXPERIMENTAL RESULT AND ANALYSIS

### A. System configurations

The SDR platform for the URLLC system consists of two GPPs and two radio transceiver devices. The UE and eNB use the device with the same configuration listed in Table II. The central processing unit (CPU) of GPP is Intel (R) Core (TM) i7-4770 with four physical cores and a maximum clock frequency of 3.4 GHz. The physical memory is 16 G and the operating system is Ubuntu14.04.

GPP is connected to a universal software radio peripheral (USRP) B210 with the frequency range from 70 MHz to 60 GHz via USB 3.0 and the maximum sampling rate of USRP B210 is 61.44 MS/s. The synchronous source OctoClock-G is used as a clock signal for the USRP B210. The SDR platform is OAI, and the carrier frequency of transmission is 2.66 GHz with 5 MHz bandwidth. The physical layer downlink is tested to evaluate the URLLC system under the frequency division duplex (FDD) and single-input single-output (SISO) model.

### B. Latency test

The downlink delay of the physical layer is tested to evaluate the latency of the URLLC system based on the OAI platform over the air. The test scenario for the system is shown in Fig. 7. Fig. 8 presents the delay of physical layer downlink and the latency is calculated according to (2) [7].

$$T_{Sum} = T_{Proc\text{-}eNB} + T_{Tx} + T_{Poc\text{-}UE} \quad (2)$$

where $T_{Sum}$ is the total delay of the physical layer. $T_{Proc\text{-}eNB}$ is the time the physical layer spends to receive the source data from the upper layer until the data are sent, including the time for encoding, rate-matching, modulation, et al., and $T_{Tx}$ is the transmission delay for transmitting the baseband data on the eNB side to the UE GPP. $T_{Proc\text{-}UE}$ is the time UE spends to get the source information from the baseband data,

TABLE I. MCS INDEX TABLE FOR PDSCH

| MCS Index | NR-TBS(Modulation) | URLLC-TBS(Modulation) |
|---|---|---|
| 0 | 848(QPSK ) | 48(QPSK ) |
| 1 | 1352(QPSK ) | 64(QPSK ) |
| 2 | 2152(QPSK ) | 72(QPSK ) |
| 3 | 3240(QPSK ) | 104(QPSK ) |
| 4 | 4224(QPSK ) | 128(QPSK ) |
| 5 | 5248(16QAM) | 160(QPSK ) |
| 6 | 6016(16QAM) | 192(QPSK ) |
| 7 | 6912(16QAM) | 256(QPSK ) |
| 8 | 7808(16QAM) | 320(QPSK ) |
| 9 | 8712(16QAM) | 432(QPSK ) |
| 10 | 9224(16QAM) | 504(QPSK ) |
| 11 | 9736(64QAM) | 640(QPSK ) |
| 12 | 11016(64QAM) | 768(QPSK ) |
| 13 | 12040(64QAM) | 888(QPSK ) |
| 14 | 13064(64QAM) | 984(QPSK ) |

TABLE II. SYSTEM CONFIGURATION

| | Parameters | Value |
|---|---|---|
| **GPP** | CPU model | Intel Core i7-4770 @ 3.4 GHz |
| | Memory | 16G |
| | Operating system | Ubuntu14.04 |
| **RF Peripheral** | Type | USRP B210 |
| | Interface | USB3.0 |
| | Frequency range | 70MHz – 6GHz |
| | Sampling rate | 61.44MHz |
| **URLLC system** | Type | OAI |
| | Carrier frequency | 2.66GHz |
| | Bandwidth | 5MHz |
| | Duplex mode | FDD |
| | Transmission model | SISO |
| **Synchronous source** | Type | OctoClock-G |

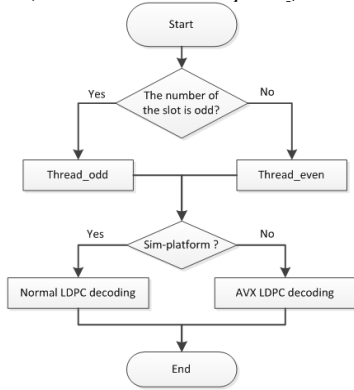

Figure 6. Part of PDSCH data process on the UE side

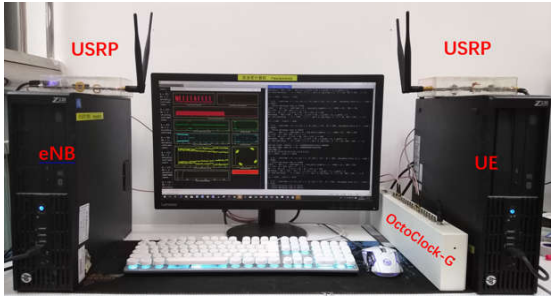

Figure 7. Experiment environment for the over-the-air platform

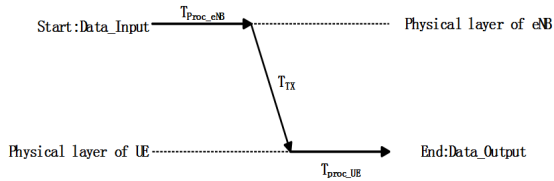

Figure 8. Downlink latency of physical layer

including the time for demodulation, de-rate matching, decoding, and other processes.

The eNB is used as the server and the UE host is time-synchronized with eNB to nanosecond level using Precision Time Protocol (PTP). The starting time $T_{Start}$ in eNB physical layer scheduling unit is processed and sent to UE. When ending time information $T_{end}$ is obtained, the latency $T_{Sum}$ of the physical layer can be calculated according to (3).

$$T_{Sum} = T_{end} - T_{Start} \quad (3)$$

The Min-Sum iterative decoding algorithm with less decoding complexity is used for LDPC decode in this system [13]. MCS from 6 to 12 are tested and the result of the physical layer delay is shown in Fig. 9. As can be seen in Fig. 9, the physical layer latency increases slowly with the MCS. The main reason is that although TBS for different MCS is different, they are of the same order of magnitude. The physical layer latency is more than 4 ms and most of the time is spent for transmission on the eNB side because it does not hand over the data to USRP B210 immediately after the baseband data are prepared.

The TBS gap between the NR system and the URLLC system in Table III gradually increases with the MCS increasing, and the gap of latency also increases between the two systems. As shown in Fig. 9, however, when the MCS add from 8 to 9, the delay gap between the two systems suddenly decreased. The NR system TBS exceeds the maximum of bits contained in a block so the multithreaded is used to deal with every block in the OAI system resulting in the decrease of delay gap as shown in Table III [13]. Then the delay gap still gradually increases as the MCS increases.

## C. Reliability test

The reliability performance is tested and evaluated based on the OAI simulation platform. LDPC is used as the coding scheme for the PDSCH, which plays an important role in improving the URLLC system reliability performance. The system also uses dual-antenna gain to improve reliability. To illustrate the reliability performance of the system, the performance under single-antenna using LDPC and Turbo is tested respectively. The performance using LDPC under single-antenna and dual-antenna is tested. By comparing the SNR-BLER diagrams of different MCS, the gain of reliability is analyzed for the scheme.

The test result of downlink performance for the LDPC scheme compared with Turbo is shown in Fig. 10. The reliability of LDPC is significantly higher than Turbo when BLER is lower. Its convergence is significantly better than Turbo, which is more obvious as the MCS decreases. The error floor problem of Turbo is more obvious than LDPC [13]. According to (2), the correct rate of 32 bytes per packet is approximately 99.999% within 1 ms. Although the maximum number of iterations for LDPC and Turbo decoding are both 6 in this experiment, the actual iteration number for LDPC decode is 2 that is more beneficial for low-latency performance than Turbo.

The test result of downlink performance for single-antenna and dual-antenna diversity with the LDPC scheme for PDSCH is shown in Fig. 11. As we can see, dual-antenna diversity is beneficial to the reliability performance of the URLLC system, and the reliability gain is more obvious with the increase of MCS. By comparing Fig. 10 and Fig. 11, it can be seen that LDPC and antenna diversity both improve the system reliability performance. The antenna diversity can provide more gain than the coding scheme for the system.

TABLE III. TIME CONSUMPTION OF LDPC DECODING

| MCS | NR System | | URLLC System | |
|---|---|---|---|---|
| | *Thread Number* | *Decode Time(ms)* | *Thread Number* | *Decode Time(ms)* |
| 4 | 1 | 0.17 | 1 | 0.0602 |
| 5 | 1 | 0.21 | 1 | 0.0610 |
| 6 | 1 | 0.25 | 1 | 0.0665 |
| 7 | 1 | 0.28 | 1 | 0.0748 |
| 8 | 1 | 0.31 | 1 | 0.0802 |
| 9 | 2 | 0.19 | 1 | 0.0849 |
| 10 | 2 | 0.20 | 1 | 0.0950 |
| 11 | 2 | 0.21 | 1 | 0.1066 |
| 12 | 2 | 0.23 | 1 | 0.1098 |
| 13 | 2 | 0.26 | 1 | 0.1159 |
| 14 | 2 | 0.28 | 1 | 0.1197 |

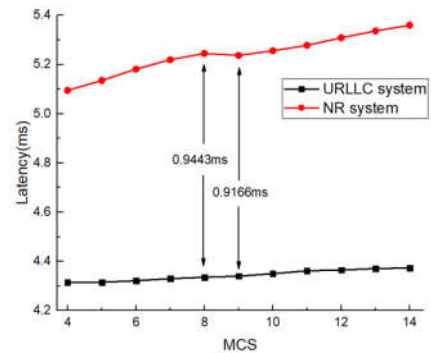

Figure 9. Test result of downlink physical layer latency

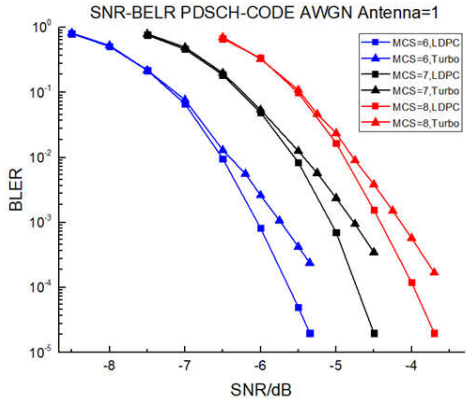

Figure 10. Reliability result for different code scheme

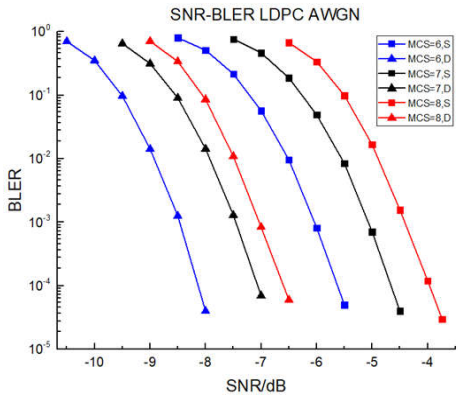

Figure 11. Reliability result for single-antenna and dual-antenna diversity

## V. CONCLUSIONS AND OUTLOOKS

The physical layer downlink of the URLLC system is designed and implemented from the aspects of delay and reliability. In terms of latency, the system physical layer downlink is based on the subcarrier spacing and the number of symbols contained in a mini-slot to meet the requirements. Although the downlink process cannot be well handled in time by either the OAI software and the USRP B210 hardware, the latency performance of the system can still achieve to the level of milliseconds, e.g., 4.3 ms in Fig. 6. From the point of reliability, the experiment uses LDPC and dual-antenna diversity as the guarantee of high-reliability. The experimental results show that the two schemes are important for system reliability. In the future, we will design more schemes from a higher layer, e.g., MAC layer, for latency and reliability optimization.

ACKNOWLEDGEMENT

This paper is supported by the National Natural Science Funding of China under Grant 61671089 and National Key Technology R&D Program of China (Grant No. 2015ZX03002009-004).